\documentclass[twocolumn]{aastex63}
\usepackage{graphicx}

\newcommand{\kms}{km~s$^{-1}$~}
\newcommand{\logZ}[1][\bfseries]{log $Z$/$Z_{\odot}$~}
\newcommand{\CII}[2][\bfseries]{C~{\sc ii}~}
\newcommand{\OI}[4][\bfseries]{O~{\sc i}~}
\newcommand{\FeII}[5][\bfseries]{Fe~{\sc ii}~}
\newcommand{\CIV}[6][\bfseries]{C~{\sc iv}~}
\newcommand{\SiII}[7][\bfseries]{Si~{\sc ii}~}
\newcommand{\SiIII}[7][\bfseries]{Si~{\sc iii}~}

\received{}
\revised{}
\accepted{}
\shorttitle{CEMP-LLS}
\shortauthors{Zou et al.}
\graphicspath{{./}{figures/}}

\begin{document}

\title{A Carbon-enhanced Lyman Limit System: Signature of the First Generation of Stars?}

\author[0000-0002-3983-6484]{Siwei Zou}
\affiliation{Kavli Institute for Astronomy and Astrophysics, Peking University, Beijing 100871, China; zousiwei@pku.edu.cn}
\affiliation{Sorbonne Universit\'e, CNRS, UMR 7095, Institut d'Astrophysique de Paris, 98 bis bd Arago, 75014 Paris, France}

\author{Patrick Petitjean}
\affiliation{Sorbonne Universit\'e, CNRS, UMR 7095, Institut d'Astrophysique de Paris, 98 bis bd Arago, 75014 Paris, France}

\author[0000-0002-5777-1629]{Pasquier Noterdaeme}
\affiliation{Sorbonne Universit\'e, CNRS, UMR 7095, Institut d'Astrophysique de Paris, 98 bis bd Arago, 75014 Paris, France}

\author[0000-0002-7864-3327]{C\'edric Ledoux}
\affiliation{European Southern Observatory, Alonso de C\'ordova 3107, Casilla 19001, Vitacura, Santiago, Chile}

\author{Raghunathan Srianand}
\affiliation{Inter-University Center for Astronomy and Astrophysics, Post Bag 4, Ganeshkhind, 411 007 Pune, India}

\author[0000-0003-4176-6486]{Linhua Jiang}
\affiliation{Kavli Institute for Astronomy and Astrophysics, Peking University, Beijing 100871, China; zousiwei@pku.edu.cn}

\author[0000-0002-4912-9388]{Jens-Kristian Krogager}
\affiliation{Sorbonne Universit\'e, CNRS, UMR 7095, Institut d'Astrophysique de Paris, 98 bis bd Arago, 75014 Paris, France}




\begin{abstract}

We present the study of a Lyman limit system (LLS) at $z_{\rm abs}$~=~1.5441 towards quasar J134122.50+185213.9 observed with VLT X-shooter. This is a very peculiar system with strong C~{\sc i} absorption seen associated with a neutral hydrogen column density of log $N$(H~{\sc i}) (cm$^{-2}$) = 18.10, too small to shield the gas from any external UV flux. The low ionization absorption lines exhibit a simple kinematic structure consistent with a single component. Using CLOUDY models to correct for ionization, we find that the ionization parameter of the gas is in the range --4.5 $<$ log $U$ $<$ --4.2 and the gas density --1.5 $<$ log~$n$(H)~(cm$^{-3}$) $<$ --1.2. The models suggest that carbon is overabundant relative to iron, [C/Fe]~$>$~+2.2 at [Fe/H] $\sim$ --1.6. Such a metal abundance pattern is reminiscent of carbon-enhanced metal-poor stars detected in the Galaxy halo. Metal enrichment by the first generation of supernovae provide a plausible explanation for the inferred abundance pattern in this system. 

\end{abstract}







\section{Introduction}

The metal content of metal-poor stars is an important clue to understanding how the first objects in the Universe were enriched and how star-formation proceeded at that time (see \citealt{fre15} for a review and references therein). Since massive metal-free stars (Population~III stars) formed from the primordial gas and then exploded as the first generation of supernovae, the gas enriched by these first supernovae will form the main source of next generation of long-lived low-mass stars, which can be detected in the halo of our Galaxy today. Thus, studies of either metal-poor stars and/or the interstellar medium (ISM) generating them are of critical importance for understanding the chemical enrichment history in the early Universe. 
 
Carbon enhancement has been detected in a large fraction of metal-poor stars, especially those at [Fe/H] $<$ --4.5 \citep{beers92,cay04,beers05,nor07,suda08,car10,fre10,an13,nord19}. These carbon-enhanced metal-poor (CEMP) stars have first been studied by \citet{bee92} who defined these stars as having [C/Fe] $>$ +1.0. Consensus on explaining metallicities in CEMP stars has not been reached yet. Several possibilities have been put forward to explain the peculiar abundance patterns in these stars \citep{car12}. One option is that they result from mass transfer from a now-extinct asymptotic giant branch (AGB) companion star \citep{her05}. In that case, the produced CEMP star exhibits enhancements in s-process neutron-capture elements during its thermally pulsating AGB phase and is thus labeled CEMP-$s$ star. This binary system model is supported by observed radial velocity variations of these stars \citep{luc05,pla13,star14}. Models have been built to interpret the abundances of CEMP-$s$ stars \citep{fuji00,iwa04,bis12}. Half of CEMP-$s$ stars also exhibit strong enhancement in r-process elements (e.g. Eu), this subset is labeled as CEMP-$r/s$ stars. Interpretation of this branch of stars is highly debatable, one assumption is that it forms out of a molecular cloud enriched by supernova ejecta \citep{bis11}. Alternatively, the carbon enhancement may be the result of the explosion of Population~III stars \citep{tom07,nom13}, that seeded (with high carbon abundance) the proto-stellar cloud of the star we see today. These are known as CEMP-no stars ([Ba/Fe] $<$0), as they show no strong enhancement in their neutron-capture elements. Perhaps the empirical evidence in support of this picture is the observation of an increasing carbon enhancement with decreasing metallicity \citep{beers05,aoki07}. Several scenarios have been suggested for the origins of CEMP-no stars \citep{nor19}, e.g. mixing and fallback in the minihalos \citep{ume03,iwa05,nom13}; spinstars \citep{mey06,mae15}; binarity \citep{suda04} and metal inhomogeneous mixing \citep{hart19}.

Detecting gas exhibiting similar abundance patterns as metal-poor stars would open a new window to investigate the imprint of the first objects in the early universe. Damped Lyman-$\alpha$ (DLA) systems, with log $N$(H~{\sc i}) $>$ 20.3 \citep{wol05}, trace most of the neutral gas in the Universe. They are considered to be intimately related to the progenitors of galaxies at high redshift. A CEMP--DLA at $z_{\rm abs}$~=~2.340 with [Fe/H]=~--3 and [C/Fe]~=~+1.53 has been first detected by \citet{cooke2010}. However, the carbon enhancement may be less than originally thought due to thermal broadening effects \citep{dutta14}. Models of metal-free nucleosynthesis can successfully reproduce the observed abundance pattern for this DLA \citep{kobayashi11}. Another DLA with [C/Fe] $>$ +0.6, [Fe/H] = --2.8 at $z$ = 3.07 has been reported by \citet{cooke12}. To date, partially ionized H~{\sc i} systems have not been detected yet with such high carbon enhancement over iron. Lyman limit systems (LLS), with 17.2$<$ log $N$(H~{\sc i}) $<$ 20.3, trace optically thick gas with complex ionization and kinematic structure in the circumgalactic medium (see \citealt{tum_ara17} and references therein). \citet{pro15} present the relative abundances of 157 high-dispersion (HD) LLSs at 1.76 $< z <$ 4.39 and tentatively conclude that the LLSs exhibit super-solar $\alpha$/Fe ratios. In contrast, their [Si/C] ratio exhibits solar relative abundance. 
However, there are handful of metal-free LLS at $z\sim 3$ \citep{fum11a} and $z=$ 4.391 \citep{rob19}, which indicates the LLSs may also be a good place to look at for the imprints of Population~III stars. Thus, highly carbon-enhanced and metal-poor LLSs at high redshift would be of particular interests for understanding peculiar nucleosynthetic processes in the early Universe. More detection of pristine gas clouds at high redshift helps to form any potential conclusion on metal-poor stars enrichment to their native ISM.  

Recently, we have detected a LLS at $z_{\rm abs}$~=~1.5441 with an enhanced carbon metallicity compared to iron, [C/Fe]~$\geq$~+2.2 at [Fe/H]~=~$-1.6$. 
In this paper, we present the observations in Section \ref{sec_measurements} and the ionization corrections in Section \ref{sec_cloudy}. We discuss the possible origin of the gas in Section \ref{sec_discussion} and give a summary in Section \ref{sec_summary}.	

\section{Observation and Data Processing}\label{sec_obs}
 
The present C~{\sc i} system, towards QSO J1341+1852, is part of the first 
sample of high redshift absorption systems selected based only on the presence of C~{\sc i} absorption. The sample contains 66 C~{\sc i} absorbers at $z>1.5$ \citep{led15}. It has been obtained by systematically searching SDSS-DR7 \citep{aba09} quasar spectra \citep{sch10} for C~{\sc i} absorption. It is complete for $W_{\rm r}$(C~{\sc i}$\lambda$1560)~$>$~0.40~\AA. Follow-up observation has been performed with the ESO Ultraviolet and Visual Echelle Spectrograph (UVES) spectrograph for 27 systems and the ESO/X-shooter spectrograph for 17 systems.
Molecular gas and its properties in the whole sample are analysed by \citet{not18}. Metallicity measurements and near-infrared (NIR) detections for X-shooter observations are presented by \citet{zou18}. 
 
Emission redshift for this QSO is 2.00, RA and DEC are 13:41:22.51 and +18:52:14.0 respectively. The $g$-band magnitude of the QSO is 17.2. The instrument X-shooter \citep{ver11} covers the full wavelength range from 300~nm to 2.5~$\mu$m at intermediate spectral resolution using three spectroscopic arms UV-blue(UVB), visible(VIS), 
and NIR. The two-dimensional (2D) and one-dimensional (1D) spectra were extracted using 
the X-shooter pipeline in its version 2.5.2 \citep{mod10}. Flux calibration has been performed using observations of standard stars provided by ESO. Generally, the final spectra are close to the nominal resolving power of R~= 4350, 7450, and 5300 in the UVB, VIS and NIR arms, for slit widths of 1.0, 0.9, and 0.9 arcsec, respectively. The converted rest-frame velocity resolutions are 69, 40 and 57 \kms respectively. Signal-to-noise (SNR) ratio per pixel achieved in the UVB, VIS and NIR arms for this object are 75, 80 and 22 respectively. 
 
\section{Measurements}\label{sec_measurements}
 
In this section, we present the observational results obtained from 
absorption lines seen in this system 
and derive the physical properties of the gas. 
In the following, all column densities $N$ are in units of cm$^{-2}$.
It must be noted that
the system is very simple with only one metal component.

\subsection{H~{\sc i}}
 
The H~{\sc i} Lyman-$\alpha$ profile together with the associated C~{\sc i} and O~{\sc i} absorption are shown in Figure \ref{fig:J1341CH}. The H~{\sc i} absorption is fitted with a Voigt profile using both the $\chi^2$ - minimization softwares VPFIT \citep{car14} and VoigtFit \citep{kro18}. The main component with log $N$(H~{\sc i}) = 18.04 $\pm$ 0.05  
is at the same redshift as the C~{\sc i} and O~{\sc i} absorptions. There is an extra weaker component present at $z$ = 1.5448 with log $N$({H~\sc i}) = 15.60 $\pm$ 0.19 for $b$ = 10 km s$^{-1}$. 

The SDSS spectrum of the quasar is presented in Figure \ref{fig:qso_attenuation}~together with a QSO spectrum template from \citep{sel16}.
The dust attenuation of the quasar is small with $A_\mathrm{V}$ and $E(B-V)$ of 0.09 and 0.033 respectively \citep{zou18}. 

\subsection{Metallicity derived from neutral oxygen}
 
We define metallicity ($Z$) of element X relative to solar as [X/H] = log $N$(X/H) $-$ log $N$(X/H)$_{\odot}$ and use solar abundances from \citet{asp_ara}. 
The oxygen metallicity, derived from 
$N$(O~{\sc i}) and $N$(H~{\sc i}), is an accurate indicator of the overall metallicity. This is because O~{\sc i} and H~{\sc i} have similar ionization potentials and are coupled by resonant charge exchange reactions, then [O/H] = [O~{\sc i}/H~{\sc i}].
For log $N$(O~{\sc i}) $<$ 14, rest-frame equivalent width $W_{\rm r}$ and column density $N$ follow a linear relation: 
\begin{equation}
N (\textrm{cm}^{-2}) = 1.13 \times 10^{20} \frac{W_{\rm r} (\textrm{\AA})}{ \lambda^2 (\textrm{\AA})f},
\end{equation}
where $f$ is the oscillator strength. The measured rest-frame equivalent width of O~{\sc i} $\lambda$1302 is $W_r = 0.08~\pm~0.01$~ \AA. With $f$(O~{\sc i} $\lambda$1302) = 0.048, 
we derive log $N$(O~{\sc i}) = 14.05 using the linear relation. The line is therefore probably saturated and this value is a lower limit.
We therefore use two methods to constrain the column density and Doppler parameter:
$a)$ we fit the $\lambda$1302 line with Voigt profiles using $b$ values 
of 5 to 10 km s$^{-1}$ ; $b)$ we use the apparent optical depth (AOD) 
method converting velocity-resolved flux into column densities 
without prior information on the curve-of-growth or the component structure \citep{fox05}. The optical depth in each velocity pixel is given by $\tau_{\rm v} = \textrm{ln}~[F_{\rm c}(v)/F(v)]$, where $F_{\rm c}$(v) is the continuum flux and $F(v)$ is the absorbed flux. 
From the results of the two methods given in Table \ref{table_J1341Z}, we conclude that log $N$(O~{\sc i}) is in the range [14.36,15.10]. 

We estimate that a robust lower limit is log $N$(O~{\sc i}) = 14.36 
(corresponding to $b$ = 8.0 km s$^{-1}$) and the lower limit on the gas metallicity is log $Z_\mathrm{lim}$ $>$ [O/H]$_\mathrm{lim}$ = $-$0.37 relative to solar. Note that this lower limit is consistent with the mean metallicity measured in other systems of the whole C~{\sc i} sample \citep{zou18}.   



\subsection{Neutral carbon} 
 
Since the ionization potential of neutral carbon (11.3 eV) is smaller than that of neutral hydrogen (13.6 eV), the detection 
of neutral carbon usually indicates the presence of neutral, cold, and well-shielded gas. 
However, at such a small $N$({H \sc i}) as what we see here, the gas cannot be shielded from the external UV flux and is probably partially ionized and warm. It is therefore very surprising to observe C~{\sc i} in this system.
This is a direct indication that the background UV flux incident on the cloud is very low and/or that the carbon metallicity is unusually high. Contrary to what would be expected from the first possibility, high-ionized ions C~{\sc iv} and Si~{\sc iv} are detected together with neutral and low-ionization species O~{\sc i}, C~{\sc ii}, Fe~{\sc ii}, Si~{\sc ii}, Si~{\sc iii}, Al~{\sc ii} and Al~{\sc iii}.

\begin{figure}
	\includegraphics[width=\columnwidth]{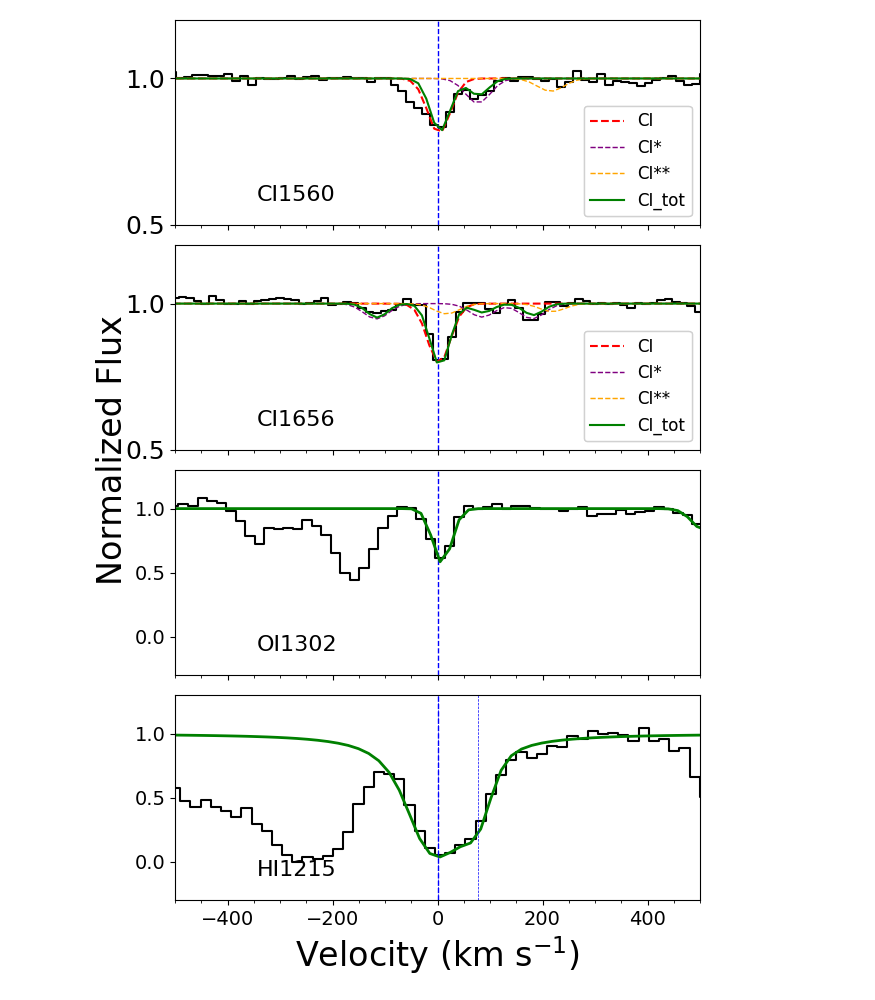}
    \caption{Voigt profile fits (green curve) to H~{\sc i}$\lambda$1215, O{\sc i}$\lambda$1302, C~{\sc i} $\lambda\lambda$1560,1656 absorption lines shown in velocity scale and centered at $z$ = 1.5441 in the J1341+1852 spectrum. Fits to the C~{\sc i} fine structure lines (C~{\sc i}* and C~{\sc i}**) are shown with purple and yellow dashed lines respectively. There is a second component of H~{\sc i} at $z= 1.5448$ indicated by the vertical dotted line.)}
   \label{fig:J1341CH}
\end{figure}

\begin{figure}
	\includegraphics[width=\columnwidth]{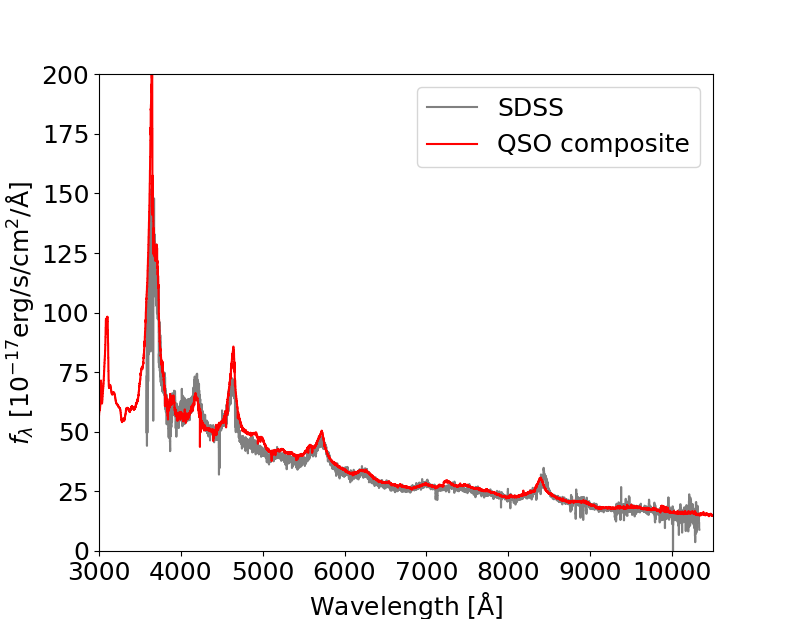}
    \caption{SDSS spectrum of J1341+1852 (in grey). The red curve is the X-shooter QSO composite template at 1~$<z<$~2 from \citet{sel16}. The template is not dust attenuated. 
    }
   \label{fig:qso_attenuation}
\end{figure}

Thanks to the reasonable resolution and good signal-to-noise ratio of the spectrum, we can disentangle 
absorption from the C~{\sc i} ground state fine-structure levels.
The fit is presented in Figure \ref{fig:J1341CH} with red, purple and yellow colors for the different levels respectively. We find that $N$(C~{\sc i}$^*$)/$N$(C~{\sc i}) ranges between 0.03 and 0.63.


\subsection{Iron} 
 
Multiple transition lines $\lambda\lambda\lambda\lambda$1608,2344,2374,2383 of Fe~{\sc ii} with very different oscillator strengths are detected. They are useful in determining the $b$ value. 
The simultaneous fit of the five lines gives
$b$ = 6.0$\pm$2.5 km s$^{-1}$ and log $N$(Fe~{\sc ii}) = 12.96$\pm$0.04. The error in $b$ is large and gives the range of possible $b$ values.
When $b$ is varied from 5 to 10 km~s$^{-1}$, 
log $N$(Fe~{\sc ii}) is well constrained within a small range [12.86,12.94]. With $b$ = 6 \kms as indicated by the Fe~{\sc ii} fit, the oxygen metallicity is $\sim$0.6~dex larger than the lower limit given above (see Table \ref{table_J1341Z}) and therefore about twice solar or larger.



\subsection{Other metal lines}
 
\begin{figure}
	\includegraphics[width=\columnwidth]{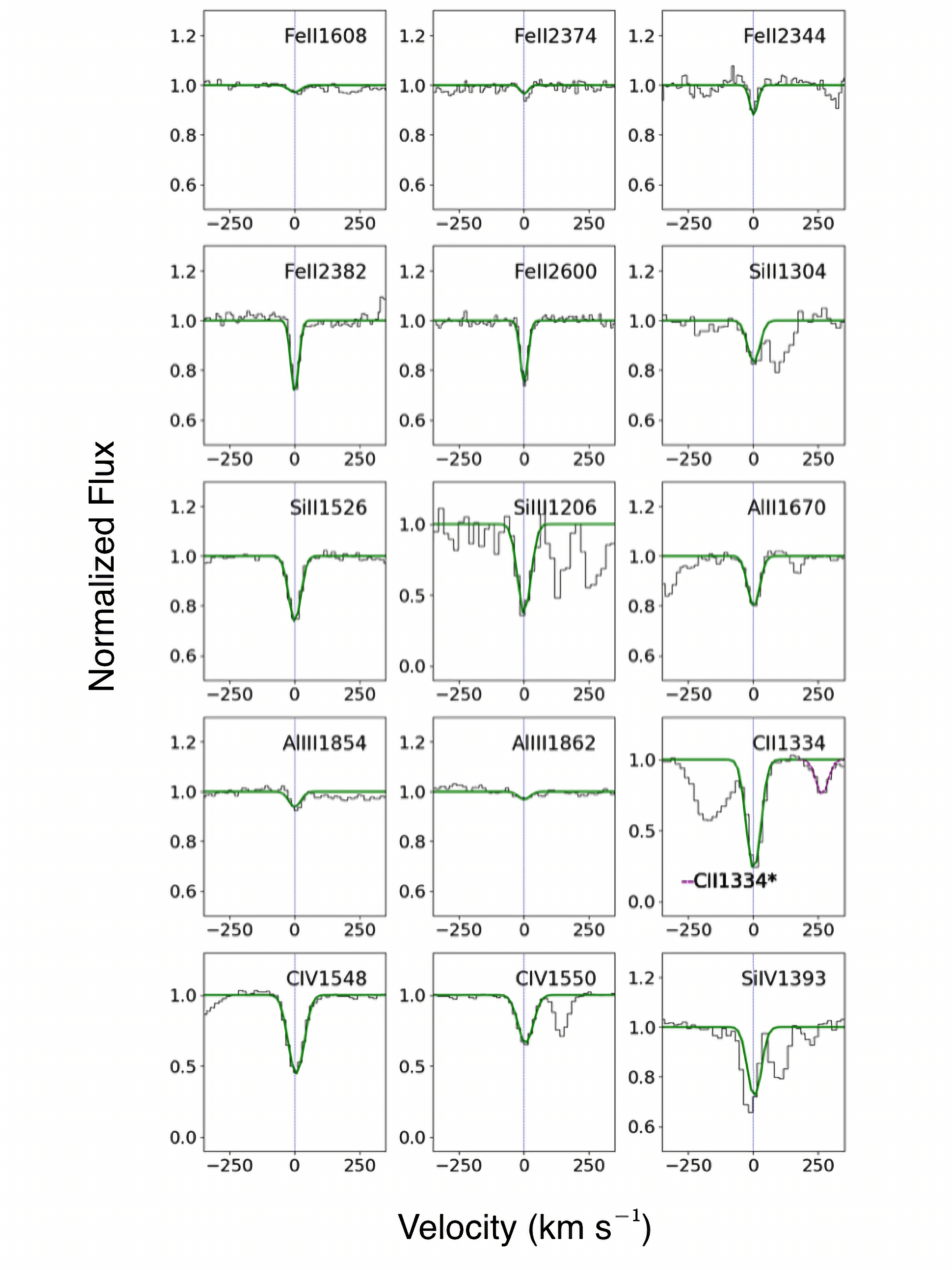}
    \caption{Voigt profile fits (green curve) of Fe~{\sc ii}, Si~{\sc ii}, Se~{}, Al~{\sc ii}, Al~{\sc iii} and C~{\sc ii} absorption lines. The Doppler parameter is $b$ = 10 km s$^{-1}$. For C~{\sc iv} and Si~{\sc iv} transition lines, $b$ = 30 km s$^{-1}$. 
    } 
   \label{fig:J1341metals}
\end{figure}

We assume the low and intermediate-ions arise from the same phase (i.e. same temperature, density and ionization parameter), so we apply the same $b$ value range ($b$ = 5, 6, 8, 10 \kms) as for Fe~{\sc ii} onto all ions except C~{\sc iv} and Si~{\sc iv}. The fit results and profiles are presented in Table \ref{table_J1341Z} and Figure \ref{fig:J1341metals}.
For Si, Si~{\sc ii}$\lambda$1304 is blended with other lines, so we only fit Si~{\sc ii}$\lambda$1526. With $b$ between 5--10 km s$^{-1}$, we find that log $N$(Si~{\sc ii}) is constrained within the range [13.62,13.87] and the ratio log~$N$(Si~{\sc iii})/$N$(Si~{\sc ii}) is in the range [0.29,1.55]. This
indicates that Si~{\sc iii} is probably the main silicon ion and confirms that
the gas is ionized. For $b$ = 5 km~s$^{-1}$, log $N$(C~{\sc ii})= 17.94$\pm$0.05.
In the following, we adopt 16.40 $\pm$ 0.24 for $b$ = 10 \kms as a robust 
lower limit of log $N$(C~{\sc ii}). However, it should be realized that if we 
were to adopt the $b$ value derived from the fit of the Fe~{\sc ii} lines, 
the carbon abundance could be an order of magnitude larger than what 
we will derive later.

In the fitting process, we find that $b$ values smaller than 10 \kms are obviously
too small to fit the high-ion C~{\sc iv} and Si~{\sc iv} absorptions, which indicates that C~{\sc iv} and Si~{\sc iv} may reside in a different phase of the same cloud. We therefore fitted absorption lines of these two ions with $b$ values of 30 km s$^{-1}$. Si~{\sc iv}$\lambda$1402 is strongly contaminated by other lines so we only fit Si~{\sc iv}$\lambda$1393. 

\subsection{Electron density}\label{sec:physical_prop}
 
We can constrain the electron density
using the observed population ratio of the 
two atomic sub-levels of the C~{\sc ii} ground state.
The lower panel in Figure 4 of \citet{sil02} gives the relation between the ratio $n$(C~{\sc ii}*)/$n$(C~{\sc ii}) and the electron density $n_{\rm e}$ under the assumption the gas is ionized and at $T$ =~10,000~K. 
Our observed log $N$(C~{\sc ii}*)/$N$(C~{\sc ii}) is in the range [$-$4.05, $-$2.69], so that we estimate the log $n_{\rm e}$ is ranging from $-$2.5 to $-$1.0. Note that we cannot use their calculation of the C~{\sc i} transitions because in our case the C~{\sc i} excitation is due to collisions with electrons and protons and not with neutral hydrogen (see their Figures 1 and 2). We will present the estimation of hydrogen density in Section \ref{sec_cloudy}.

\begin{center}
\begin{table*}
\small
  \caption{Column densities (log $N$) derived for different assumed Doppler parameter values ($b$) in the $z$~=~1.5441 absorption system toward J1341+1852. Both results from Voigt profile fits and AOD method are presented. 
  \label{table_J1341Z} }
\centering
\begin{tabular}{lccccc}
\hline
\multicolumn{6}{c}{ $z$ (C~{\sc i}) =1.5441}\\
\hline
\multicolumn{5}{c}{  log $N$ (VPFIT) } & log $N$ (AOD)  \\
\hline
Species &  $b$=5.0 km s$^{-1}$& $b$=6.0 km s$^{-1}$ & $b$=8.0 km s$^{-1}$& $b$=10.0 km s$^{-1}$ &   \\
 C {\sc i}      
 & 13.67$\pm$0.06 & 13.53$\pm$0.06 &  13.47$\pm$0.10 & 13.44$\pm$0.10 & 13.21$\pm$0.01 \\ 
 C {\sc i}*     
 & 13.26$\pm$0.07 & 13.25$\pm$0.08 &  13.25$\pm$0.08 & 13.24$\pm$0.07 &                \\
 C {\sc i}**    
 & $<$12.54 & $<$12.25 &  $<$12.51 & $<$12.71 &                \\ 
 C {\sc ii}     
 & 17.90$\pm$0.05 & 17.58$\pm$0.18 &  17.56$\pm$0.19 & 16.40$\pm$0.24 & $>14.35$       \\  
 C {\sc ii}*    
 & 13.85$\pm$0.12 & 13.74$\pm$0.10 &  13.72$\pm$0.09  & 13.71$\pm$0.07 &                \\
 Fe {\sc ii}    
 & 12.94$\pm$0.04 & 12.96$\pm$0.04 &  12.86$\pm$0.05 & 12.82$\pm$0.05 &  12.76$\pm$0.01 \\
 Mg {\sc ii}    
 & 15.52$\pm$0.18 & 14.42$\pm$0.24 &  13.58$\pm$0.05 & 12.90$\pm$0.07 & 11.42$\pm$0.03 \\  
 Al {\sc ii}    
 & 12.45$\pm$0.07 & 12.30$\pm$0.04 &  12.30$\pm$0.05 & 12.28$\pm$0.05 & 12.20$\pm$0.06 \\
 Si {\sc ii}    
 & 13.87$\pm$0.10 & 13.64$\pm$0.04 &  13.62$\pm$0.05 & 13.60$\pm$0.05 & 13.50$\pm$0.08 \\  
 Al {\sc iii}   
 & 12.16$\pm$0.10 & 12.15$\pm$0.08 &  12.15$\pm$0.09 & 12.09$\pm$0.09  & 12.08$\pm$0.20 \\
 Si {\sc iii}   
 & 15.42$\pm$0.55 & 14.65$\pm$0.79 &  13.91$\pm$0.75 & 13.41$\pm$0.85  & 12.76$\pm$0.10 \\
 O {\sc i}      
 & 15.10$\pm$0.39 & 14.60$\pm$0.16 &  14.36$\pm$0.23 & 14.40$\pm$0.10  & 14.39$\pm$0.01 \\ 
 H {\sc i}      
 & 18.12$\pm$0.04 & 18.10$\pm$0.06 &  18.04$\pm$0.05 & 18.09$\pm$0.05  & 18.00$\pm$0.01 \\
 H {\sc i}$^{a}$ 
 & 15.62$\pm$0.58 & 15.60$\pm$0.25 &  15.60$\pm$0.19 & 15.70$\pm$0.20  &  \\
 \hline
 \multicolumn{3}{c}{ $b$ = 30.0 km s$^{-1}$} & \\
 \hline
 C {\sc iv}& \multicolumn{3}{l}{  13.89 $\pm$ 0.01 }& \\
 Si {\sc iv} & \multicolumn{3}{l}{  $<$ 13.05 }&\\
 \hline 
 \multicolumn{5}{l}{a. The second component of H~{\sc i} is at $z$ = 1.5448.  }  \\
\end{tabular}
\end{table*}   
\end{center}

\section{Ionization correction}\label{sec_cloudy}
 
In this section, we will model the ionization state of the gas in the current system. 
Indeed, for systems with log $N$(H~{\sc i}) $<$ 19.3, ionization corrections can be important
and metallicities cannot be derived directly using column densities of only neutral or singly-ionized species.
This is contrary to what can be done in the case of DLAs. For this we use the photoionization code CLOUDY \citep{fer17} to model the ionization state of the gas.
It is important to note that for low and intermediate ionization species, only one narrow absorption component is seen.
Given this simple kinematic structure of this system, we assume that the system is a single cloud.
\begin{table*}
\small
  \caption{Parameters and outputs of photoionization models for the system at $z_{\rm abs}$~=~1.5441 toward J1341+1852. 
  The upper part of the Table shows the parameter ranges and steps used in the model. The lower part gives the results (column density, metallicities relative to solar and metallicity ratios) for the best-fit.
  \label{table_model}}
\centering
\begin{tabular}{llllr}
\hline

UV source  & Parameters         & min   & max       & Best Fit\\
KS19       & log $n$(H) (cm$^{-3}$) & -4.00 & 0.00      & $-$1.5 $\sim$ $-$1.2   \\
           & log $U$            & -1.99  & -5.99    & $-$4.5 $\sim$ $-$4.2   \\
           & log $n_{\rm e}$ (cm$^{-3}$) & -4.00  & 0.00     & $-$1.5 $\sim$ $-$1.2   \\
           & $T$ (K)              & 1,000 & 10,000    & 9,000 $\sim$ 9500  \\
  \hline
Best-fit   &[C~{\sc i}/C~{\sc ii}]      &&& --2.70 \\
           &[C~{\sc ii}/C~{\sc iv}]     &&& 2.05   \\
           &[Fe~{\sc ii}/Fe~{\sc iii}]  &&& --0.28 \\
           &[Si~{\sc iii}/Si~{\sc ii}]  &&& --0.10 \\
           &[Si~{\sc iii}/Si~{\sc iv}]   &&& 0.58 \\
           
           &[Fe/H]                   &&& --1.6 \\           
           &[O/H]                    &&& 0.0  \\
           &[C/H]                    &&& 0.6  \\  
           &[Si/H]                   &&& --1.3 \\
           &[Al/H]                   &&& --1.5 \\
           &[C/O]                    &&& 0.6  \\
           &[C/Fe]                   &&& 2.2  \\
           &[Si/Fe]                  &&& 0.4  \\
           &[Al/Fe]                  &&& 0.1  \\             

\hline   
\end{tabular}
\end{table*}

\begin{figure}
	\includegraphics[width=\columnwidth]{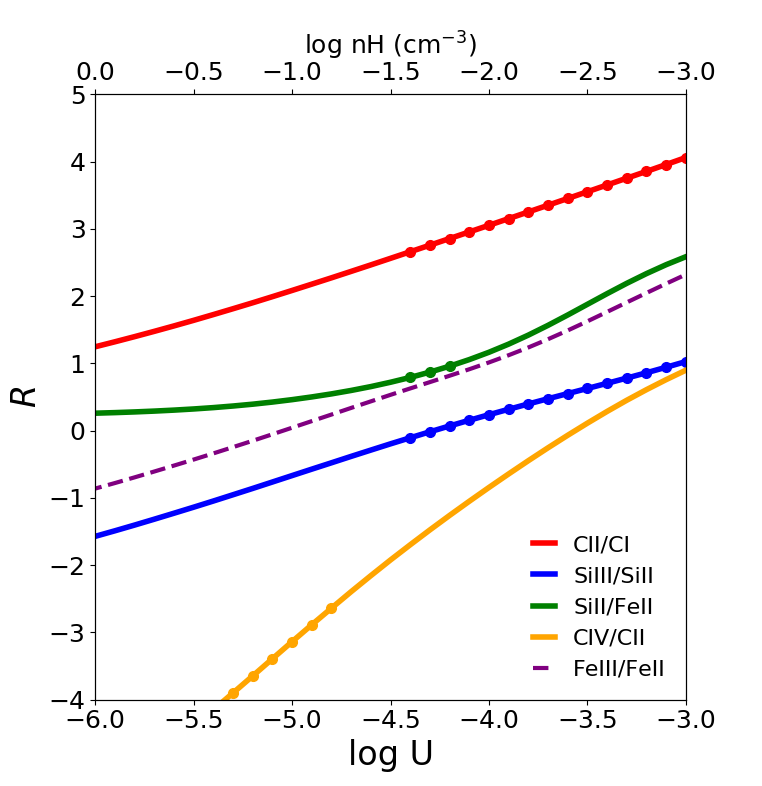}
    \caption{Column density ratios 
    from CLOUDY models against ionization parameter log $U$. Densities corresponding to $U$ are given in the top axis. The ratios are plotted on a logarithmic scale. The ionizing flux is the metagalactic UV background from \citet{kha19}, abundance pattern is solar and overall metallicity is \logZ = 0.3. Solid curves correspond to models, dots represent the observationally-allowed values.}
   \label{fig:bg_ratio}
\end{figure}

We firstly assume that the system is photo-ionized by the metagalactic ionizing UV background from \citet{kha19} only. We add the cosmic microwave background (CMB) at $z$ = 1.5441. Then we vary the hydrogen volume density log~$n$(H) (cm$^{-3}$) within the range [$-$4,0], with the two stopping criteria, log $N$(H~{\sc i}) = 18 and $T$ = 100 K. Ionization state of the gas is described by the ionization parameter $U$ = $\frac{\Phi (H)}{n(H)c}$, where $\Phi(H)$ is the flux of ionizing photons. Therefore the ionization parameter log $U$ is in the range  [$-$1.99,$-$5.99]. In Figure \ref{fig:bg_ratio}, we plot the logarithmic column density ratios $R$ of different ions as a function of log $U$. Acceptable ranges derived from observations are indicated by colored points. We can tell that the ratios of low- and intermediate-ions can be reproduced within the range log $U$ = [$-$4.5,$-$4.2]. We also find that if relative solar metallicity is assumed, the column density ratios do not depend on the exact overall metallicity. With the metallicity varying between --2 $<$ \logZ $<$ 1, the ratios log $N$(C~{\sc iii})/$N$(C~{\sc ii}) and log $N$(Si~{\sc iii})/$N$(Si~{\sc ii}) barely change. The ratio of log $N$(Si~{\sc ii})/$N$(Fe~{\sc ii}) is slightly higher at lower metallicity. 

However, 
we find that with the same ionization parameter range (i.e. gas density range), the absolute column densities  cannot be produced.
At $-$4.5 $<$ log $U$ $\sim$ $<$ --4.2, we can reproduce $N$(C~{\sc i}) and $N$(O~{\sc i}) with \logZ = 0.3 and $N$(C~{\sc ii}) with at least \logZ = 0.6. But the calculated log $N$(Fe~{\sc ii}) at \logZ = 0.6 is in the range of [13.71,14.90], which is much larger than the observed upper limit log $N$(Fe~{\sc ii}) = 12.94. To reproduce the absolute observed Fe~{\sc ii} column density, we have to decrease the iron metallicity to \logZ $\leq$ $-$1.4. 
This metallicity discrepancy between iron and carbon implies that the relative abundance of C over Fe should be at least 2.2 dex higher than the solar value. 
In addition, the
relative C to O abundance should be larger than solar.

\begin{figure}
	\includegraphics[width=\columnwidth]{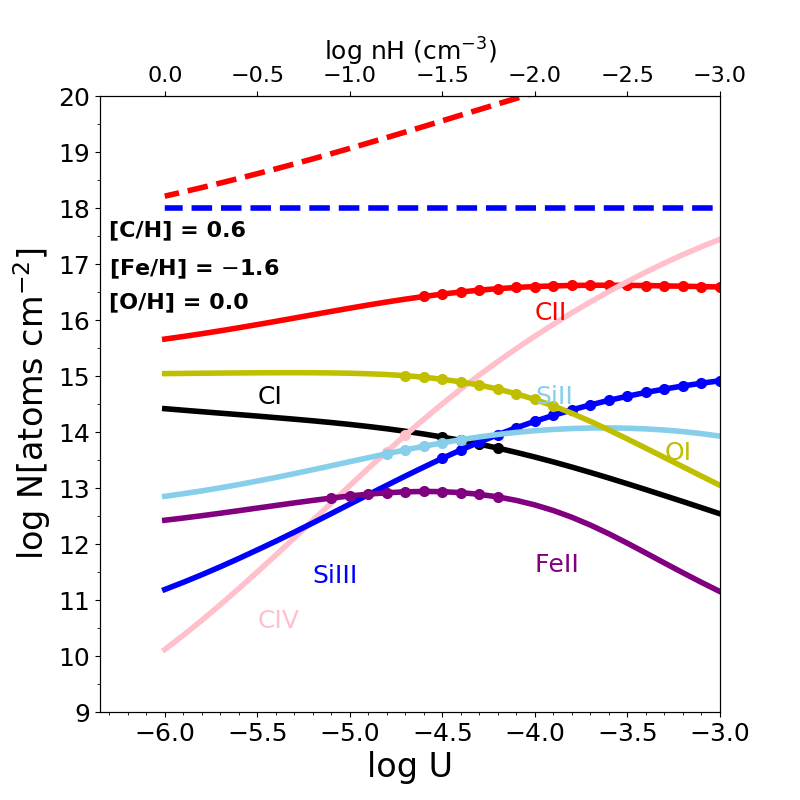}
    \caption{Ionization state modeling with UV metagalactic background from \citet{kha19}, abundance pattern is settled as non-solar with [C/H] = 0.6 and [Fe/H] = --1.6. The best fit parameters and results are listed in Table \ref{table_model}. The solid curves are from models, dots are the range of values allowed by the observations.}
   \label{fig:bg_Z03_ism}
\end{figure}

Therefore, as Figure \ref{fig:bg_Z03_ism} shows, we have to adjust the carbon abundance to 4 times higher than solar, Fe abundance to 0.02 times solar, Si abundance to 0.04 times solar and Al abundance to 0.03 solar. We then find that with $-4.5<$ log $U$ $<-$4.2, nearly all the column densities can be reproduced. The calculation parameter ranges and best-fit results are presented in Table \ref{table_model}. The optimal temperature range is between 9,000 -- 9,500~K while we fix the density and metal abundances. 
We therefore conclude that {\bf [C/Fe] = +2.2, [C/O] = +0.6 at [Fe/H] = $-$1.6}.

\vspace{1cm}

\section{Discussion}\label{sec_discussion}
To summarize the CLOUDY models above, if we only consider the metagalactic background as the ionization source, we can reproduce the observed column densities with an ionization parameter in the range $-$4.5 $<$ log $U$ $<$ $-$4.2. 
The model also indicates that log $n_{\rm e}$ = log $n$(H) and should be close to $-1.5$
which is consistent with our estimation in Section \ref{sec:physical_prop}. Of course, the UV flux can be stronger due to the possible presence of nearby sources. In that case, the density would be larger. A factor of three larger would be acceptable.
Column densities of C~{\sc ii}, O~{\sc i} and Fe~{\sc ii}  cannot be reproduced with the same ionization parameter unless relative abundances are different from solar and large carbon enhancement is needed. 

Given that the gas exhibits high carbon enhancement, [C/Fe] = +2.2 at [Fe/H] = --1.6, we conjecture that the gas shares abundances similarity with CEMP stars. We discuss the possible scenarios that would yield such metal abundances and explain the origin of the gas in the following.

\subsection{Depletion and extinction}
Dust depletion is usually invoked to explain iron deficiency
relative to carbon. We have shown however that the gas is highly ionized and
warm. Although some depletion is likely, it cannot be as large as
to explain what we observe. For this, the gas should be dense and cold
as in our Galaxy. This is unlikely to be the case given the
low H~{\sc i} column density and the presence at the same time of
highly ionized species. We should recall as well that we have used
the lower limit for $N$(C~{\sc ii}) and this column density is
most probably larger. Furthermore, \citet{fum16} constructed ionization models to investigate the properties of the HD-LLSs in \citet{pro15} and 77 LLSs from literature, they show that the LLSs at 1.8 $\leq z \leq$ 4.4 are ionized and typically reside in environments with low dust content. 


\subsection{Scenario for the gas phase abundances}
The most promising origin of the carbon enhancement is the production of metals in a
core-collapse supernovae which undergoes mixing and fallback processes \citep{ume03,tom07,heg10,cooke14,tom14,ish14}. In this scenario, the explosion energy, progenitor mass, metallicity, and gravitational energy \citep{nom13} are the key parameters affecting the final metal abundance pattern. When the explosion energy is not high enough to expel Fe--peak elements from the innermost layers,  
a large fraction of these elements eventually fall back onto the remnant. 
Therefore the interstellar gas polluted by the ejected gas will show 
some C and O enhancement with respect to Fe. This process leads to the generation of CEMP stars. Smaller explosion energy leads to larger fallback, thus with the same progenitor mass, higher [C/Fe]. 

\citet{ume03} proposed that the core-collapse supernovae with a progenitor of mass 20--130 $M_{\odot}$ can explain the abundances observed in C-rich stars. 
The hyper metal-poor star (HMP) 
HE 0107--5240 has relative abundances similar to what we observed \citep{col06}. 
Different progenitor masses, metallicity and explosion energy were adopted to explain the abundances of this HMP star. Most of them adopted $M =25 M_{\odot}$ as the progenitor mass and different explosion energy ($E_{51} = E/10^{51}$ erg): 0.3 \citep{ume03}, 0.71 \citep{iwa05}, 1 \citep{ish14} and 5 \citep{tom14}. In addition, as shown in Figure 8 of \citet{nom13}, the faint supernovae nucleosynthetic yields also match our carbon abundance relative to Fe ([C/Fe] = +2.2). Similar metal relative abundances of our gas have also been detected in CEMP stars HE 2139--5432, HE 1150--0428, and G 77--61 \citep{tom14}. $E_{51}$ = 5 is set for the previous two stars and $E_{51}$  = 1 for the latter in the models in \citet{tom14}. We therefore propose that our gas was enriched by core-collapse supernovae with $>20M_{\odot}$ progenitor mass and a small explosion energy. The deficiency of Al and Si with respect to C and O that we observe is also expected and further suggests that the gas has been enriched by faint supernovae. 

It would be interesting to estimate the number of supernovae needed to enrich the gas we see to such metallicities. However, we have no information about the geometry of the cloud and especially the extension perpendicular to the line of sight. It seems probable however that several supernovae are needed. The region we observe is therefore probably peculiar in terms of environment. 

We propose that the mixing and fallback scenario as the most promising origin of the carbon enhancement in the gas, however, other channels are not ruled out. The mixing and fallback scenario is applied to the explanation of CEMP-no stars, however, CEMP-no stars are mainly found at [Fe/H] $<$ --2. Thus, to explain our [Fe/H], an extra Fe contribution from a long-lived SN Ia is maybe a solution. Alternatively, low-mass (1--4 $M_\odot$) AGB stars can provide carbon enhancement to the local ISM \citep{kar10} 0.15 Gyr after the formation of the system \citep{kobayashi11b}. However, as shown in the yields by \citet{kobayashi11b}, only low-mass AGB stars without faint supernovae cannot explain our [C/Fe] at $z = 1.5441$. Instead, our [Fe/H] may indicate the signature of CEMP-$r/s$ stars (e.g. BS 16080--175 with [C/Fe]  = +1.8 at [Fe/H] = --1.9, \citealt{allen12}). One possible scenario for this subset of CEMP stars is that they are formed in a binary system. The gas clouds, from which the binary system was born, was $r$-rich because of supernova explosions of previous-generation stars \citep{aba16}.

Nevertheless, with the present relative abundances of only a few elements along the line-of-sight, we cannot determine that the gas shares similar features with CEMP-no, CEMP-$s$, or CEMP-$r/s$ stars, or fine-tune the actual enrichment history of this peculiar system. It would thus be of particular interest to obtain higher spectral resolution data to further investigate our conjecture and constrain better the physical state of this very peculiar gas.

\section{Summary}\label{sec_summary}
 
We have presented a detailed analysis of a very peculiar LLS at $z_{\rm abs}$ = 1.5441 towards quasar J134122.50+185213.9.

1. Using X-Shooter data, we surprisingly detect C~{\sc i} absorption lines when the neutral hydrogen column density is small, log $N$(H~{\sc i}) = 18.04$\pm$0.05. 
We derive a gas-phase overall metallicity lower limit of [O /H ] = $-$0.37. 

2. The kinematics of the system is simple
with only one absorption component detected for neutral and singly ionized species.

3. We calculate the ionization correction with CLOUDY models. The gas is ionized with  
an ionization parameter of log $U$ = $-$4.5
and a density log~$n$(H)~$\sim -$ 1.5 (cm$^{-3}$). 
We show that oxygen is roughly solar, 
[Fe/H] $\sim$ $-$1.6, and [C/Fe] = +2.2. 

We suggest that the gas has been polluted by
supernovae with $M > 20 M_{\odot}$, 
which have undergone a mixing fallback process.

4. It would be very difficult to avoid
a carbon metallicity enhancement for this system. 
However, to derive the accurate metal abundances and the exact physical state of the gas and therefore
settle the origin of this unusual set of metallicities,
high-resolution spectroscopic data are crucially needed.





.



\label{lastpage}

\section*{Acknowledgements}

We thank the anonymous referee for constructive comments and suggestions. SZ and LJ gratefully acknowledge support from the National Science Foundation of China (11721303, 11890693) and the National Key R\&D Program of China (2016YFA0400703). PN, RS, and PPJ gratefully acknowledge support from the Indo-French Centre for the Promotion of Advanced Research (Centre Franco-Indien pour la Promotion de la Recherche Avanc\'ee) under contract No. 5504-B. PN and JKK acknowledge support from the French Agence Nationale de la Recherche under grant no ANR-17-CE31-0011-01 (project “HIH2” – PI Noterdaeme). We thank fruitful discussions with Sergei Balashev, Xiaoting Fu, Tilman Hartwig, Kohei Inayoshi and Chiaki Kobayashi, and valuable suggestions on the early version of manuscript by Zheng Cai and Ravi Joshi. We thank Sebastian Lopez and Thomas Kr\"uhler for help with the X-shooter data analysis.
 



\bibliography{phd}{}
\bibliographystyle{aasjournal}






\label{lastpage}

\end{document}